\documentclass[journal]{IEEEtran}

\ifCLASSINFOpdf
\else
   \usepackage[dvips]{graphicx}
\fi
\usepackage{url}
\usepackage{amsmath}
\usepackage{cite}
\usepackage{bbm}
\usepackage{relsize}

\usepackage{booktabs}
\usepackage{array, multirow}

\usepackage{graphicx}

\begin{document}

\title{Generalising AUC Optimisation to Multiclass Classification for Audio Segmentation with Limited Training Data}

\author{Pablo Gimeno, Victoria Mingote, Alfonso Ortega, Antonio Miguel and Eduardo Lleida
\thanks{This project has received funding from the European Union's Horizon 2020 research and innovation programme under the Marie Sk\l{}odowska-Curie grant agreement No 101007666, by FEDER/Spanish Ministry of Economy and Competitiveness (TIN2017-85854-C4-1-R) and the Government of Arag\'{o}n (Reference Group T36\_20R).}
\thanks{The authors are with ViVoLab research group from the Arag\'{o}n Institute for Engineering Research (I3A), University of Zaragoza, 50018 Spain (e-mail: pablogj@unizar.es).}
}

\markboth{IEEE SIGNAL PROCESSING LETTERS}
{Shell \MakeLowercase{\textit{et al.}}: Bare Demo of IEEEtran.cls for IEEE Journals}
\maketitle

\begin{abstract}
Area under the ROC curve (AUC) optimisation techniques developed for neural networks have recently demonstrated their capabilities in different audio and speech related tasks. However, due to its intrinsic nature, AUC optimisation has focused only on binary tasks so far. In this paper, we introduce an extension to the AUC optimisation framework so that it can be easily applied to an arbitrary number of classes, aiming to overcome the issues derived from training data limitations in deep learning solutions. Building upon the multiclass definitions of the AUC metric found in the literature, we define two new training objectives using a one-versus-one and a one-versus-rest approach. In order to demonstrate its potential, we apply them in an audio segmentation task with limited training data that aims to differentiate 3 classes: foreground music, background music and no music. Experimental results show that our proposal can improve the performance of audio segmentation systems significantly compared to traditional training criteria such as cross entropy.
\end{abstract}

\begin{IEEEkeywords}
audio segmentation, limited training data, multiclass AUC optimisation
\end{IEEEkeywords}

\IEEEpeerreviewmaketitle

\section{Introduction}
\IEEEPARstart{A}{udio} segmentation aims to obtain a set of labels so that an audio signal can be separated into homogeneous regions and then classify those regions into a predefined set of classes, e.g., speech, music or noise. Due to the large increase in multimedia content generation in the last few years, the relevance of these systems, that can index and analyse multimedia streams, is becoming more and more significant.

In this context, we introduce music-related audio segmentation systems, that deal with the musical content of audio signals. Currently, many systems focus on the separation of speech and music \cite{speech_music1,speech_music2}, and the music detection task \cite{md1}. Both tasks are especially relevant in the broadcast content in order to monitor copyright infringements \cite{copyright} and in the context of document information retrieval\cite{dir1,dir2}. In this paper, we concentrate our attention on the relative music loudness estimation task, whose goal is to separate an audio signal into three classes: foreground music, background music and no music. This task was proposed initially in the MIREX \cite{mirex} 2018 and 2019 technological evaluations, and it is supported by the release of the OpenBMAT dataset \cite{openbmat} for training and evaluation, which we use in this work.

An important part of audio segmentation systems is how they can be properly evaluated. Some binary segmentation systems rely on the receiver operating characteristic (ROC) curve as a method to represent its performance. This curve plots false positive rates (FPR) versus true positive rates (TPR) for all the possible detection thresholds. Moreover, the area under the ROC curve (AUC) is a performance measurement for classification problems that provides an aggregate value which considers all the possible detection thresholds. Several works have already proposed to directly optimise the AUC for different applications with promising results \cite{AUC_old,JFA_auc}. To the best of our knowledge, up until now all the presented works that aim to optimise the AUC using a deep learning framework are limited to binary tasks, as this is the usual situation when applying the AUC metric. Furthermore, it should also be noted that the performance of most deep learning solutions is strongly dependent on the data they were trained on. Our most recent work \cite{auc_music_detection} has shown how AUC optimisation can outperform traditional training objectives in the context of limited training data.

Therefore, the major contribution of this work is twofold. Firstly, we aim to generalise the AUC optimisation techniques to an arbitrary number of classes, taking as starting point the multiclass variations of the AUC metric proposed in the literature. Secondly, we aim to overcome the issues derived from a limited training data scenario by introducing these recently presented multiclass AUC optimisation techniques.

The remainder of the paper is organised as follows: Section
\ref{sec:beyond} introduces the multiclass AUC metrics and the proposed multiclass AUC training objectives. Section \ref{sec:exp_setup} presents the experimental setup, describing the neural network architecture, the dataset considered and the metrics used in the evaluation. In Section \ref{sec:res}, we describe the results obtained for our audio segmentation system. Finally, a summary and the conclusions are presented in Section \ref{sec:conc}.
\section{Beyond binary AUC optimisation}
\label{sec:beyond}
\subsection{From binary to multiclass AUC}
\label{sec:mc_AUC}
Due to its intrinsic nature, ROC analysis \cite{roc_analysis} and the AUC metric are commonly used in binary tasks, where the goal is to separate two different sets: targets and non-targets. However, different studies have proposed an extension of the AUC metric in order to be used as a performance measure in tasks with more than two classes \cite{multiclass_roc_analysis,AUC_ovo,auc_mu,AUC_ovr}. Generally, two main approaches are known:

\begin{itemize}
    \item \textbf{one-versus-one (OVO) approach}: originally proposed in \cite{AUC_ovo}, this approach extends the AUC metric to the multiclass domain by averaging pairwise comparisons of classes. Supposing a problem with $c$ classes ($c > 2$), we define the set $C_i$ as the set of elements belonging to class $i$, and $\text{AUC}(C_i,C_j)$ as the binary AUC metric obtained using the elements belonging to class $i$ as targets and the elements belonging to class $j$ as non-targets. Therefore, the overall multiclass AUC can be defined according to the following equation:
    \begin{equation}
        \text{AUC}_{\text{OVO}} = \frac{2}{c (c-1)}  \sum_{i=0}^{c-2} \sum_{j=i+1}^{c-1} \widehat{\text{AUC}}(C_i,C_j)\,.
        \label{eq:auc_ovo}
    \end{equation}
    In general, for more than two classes, $\text{AUC}(C_i,C_j) \neq \text{AUC}(C_j,C_i)$. This problem is tackled in \cite{AUC_ovo} by using the modified version $\widehat{\text{AUC}}(C_i,C_j)$ as the measure of separability between classes $i$ and $j$. This is done according to the following equation:
    \begin{equation}
        \widehat{\text{AUC}}(C_i,C_j) = \frac{1}{2} [\text{AUC}(C_i,C_j) + \text{AUC}(C_j,C_i)]\,.
        \label{eq:auc_ovo_aux}
    \end{equation}
    
    \item \textbf{one-versus-rest (OVR) approach}: This approach was firstly proposed in \cite{AUC_ovr}, and is based on the idea that a classification system with $c$ classes ($c > 2$) can also be interpreted as a system running $c$ binary classifiers in parallel. With this idea in mind, the multiclass AUC metric can be defined as the mean of the $c$ AUC binary metrics obtained, as expressed in the following equation:
    \begin{equation}
        \text{AUC}_{\text{OVR}} = \frac{1}{c}  \sum_{i=0}^{c-1}{\text{AUC}}(C_i, C_i^C)\,,
        \label{eq:auc_ovr}
    \end{equation}
    where $C_i^C$ is the complement set of $C_i$ and $\text{AUC}(C_i, C_i^C)$ is the AUC metric obtained setting the elements belonging to class $i$ as targets, and the elements belonging to the other classes as non-targets.
\end{itemize}

So far, different binary audio related tasks have demonstrated the capabilities of the AUC and partial AUC optimisation techniques in a deep learning framework: speaker verification systems have been trained by optimising the AUC \cite{mingoteAUC} and partial AUC \cite{bai2020speaker,bai2020partial} metrics. Authors in \cite{vad_auc} use the same training objective to develop a speech activity detection system. An equivalent approach is used in \cite{ge2e} considering tuples of positives and negatives examples for speaker verification. An evaluation of the most popular loss functions based on the metric learning paradigm is presented in \cite{metric_learning} for the speaker recognition task. Our most recent work \cite{auc_music_detection} demonstrated the feasibility of partial AUC optimisation techniques in the music detection task, suggesting that it can improve the system performance when compared to traditional training criterion in a limited training data scenario. 

Building upon the description of the multiclass AUC metrics shown, in this work we aim to extend the AUC optimisation techniques to the multiclass classification framework in order to provide a generalisation that can be applied to an arbitrary number of classes. To the best of our knowledge, this is the first approach to multiclass AUC optimisation.

\subsection{Multiclass AUC optimisation}
\label{sec:mAUC_opt}
 As the starting point of the formulation for the multiclass AUC loss functions, we need to refer to the already proposed binary AUC optimisation definition. A more detailed description can be found in \cite{auc_music_detection}. Consider a dataset $\Pi \! = \! \{\mathbf{X},\mathbf{Y}\}$ where $\mathbf{X} \! = \! \{ \mathbf{x_1} , ...\, , \mathbf{x_N}\}$ is the acoustic features set with $N$ different examples, and $\mathbf{Y} \! = \! \{y_1, ...\, , y_N\}$ are the binary labels defining each of the elements in $\mathbf{X}$ as target or non-target examples (1 or 0 respectively). The neural network can be expressed then as a function $f_\theta: {\rm I\!R}^D \to {\rm I\!R}$ depending on a set of parameters $\theta$ and mapping the input space of dimension $D$ to a real number representing the target score. Additionally, we define sets $\mathbf{S^+} \! \! = \! \{f_{\theta}(\mathbf{x_i})\: \forall \mathbf{x_i} \in \mathbf{X}\, |\, y_i = 1\}$, which is the set of neural network scores for the target examples in $\mathbf{X}$, and $\mathbf{S^-} \! \! = \! \{f_{\theta}(\mathbf{x_i})\: \forall \mathbf{x_i} \in \mathbf{X}\, |\, y_i = 0\}$ that represents the set of neural network scores for the non-target examples in~$\mathbf{X}$. Cardinalities of those sets are $N^+$ and $N^-$ respectively. In batch training, the subsets $\mathbf{S^+}$ and $\mathbf{S^-}$ are constrained to the positive and negative examples present in the current minibatch. Then, the binary AUC loss can be defined as

\begin{equation}
    \text{AUC} = \frac{1}{N^{+}N^{-}} \sum_{i=1}^{N^{+}}\sum_{j=1}^{N^{-}} \mathlarger{\mathbbm{1}} \Big( s_i^+ > s_j^-\Big)\,,
    \label{eq:auc_loss}
\end{equation}
where $\mathlarger{\mathbbm{1}}(\cdot)$ is equal to `1' whenever $s_i^+ > s_j^-$ and `0' otherwise. This expression can be rewritten using the unit step function as
\begin{equation}
    \text{AUC} = \frac{1}{N^{+}N^{-}} \sum_{i=1}^{N^{+}}\sum_{j=1}^{N^{-}} u \Big( s_i^+ - s_j^-\Big)\,.
    \label{eq:auc_loss_step}
\end{equation}
 
Now, we apply the sigmoid approximation described in \cite{mingoteAUC} to overcome the non-differentiability issues derived from the unit step function, allowing the backpropagation of gradients. This leads to the expression of the aAUC (where `\textit{a}' stands for approximated) described in the following equation:
\begin{equation}
   \text{aAUC} = \frac{1}{N^{+}N^{-}} \sum_{i=1}^{N^{+}}\sum_{j=1}^{N^{-}} \sigma \Big(\delta \big(s_i^+ - s_j^-\big)\Big)\,,
    \label{eq:aAUC_loss}
\end{equation}
where $\sigma(\cdot)$ is the sigmoid function and $\delta$ is a hyperparameter that controls the slope of the sigmoid.

As it has already been explained in section \ref{sec:mc_AUC}, the multiclass variations of the AUC metric are built upon the computation of several binary AUC metrics, so Eq. (\ref{eq:aAUC_loss}) can be used as the main element to build our proposed multiclass AUC training objectives. Just by modifying which classes are used as targets and non-targets we can derive both multiclass loss functions. In the case of the OVO approach, through the sigmoid approximation and by combining Eq. (\ref{eq:auc_ovo}) and (\ref{eq:auc_ovo_aux}), it is straightforward to obtain the expression for the $\text{aAUC}_{\text{OVO}}$ loss. Similarly, for the the OVR approach the expression for the $\text{aAUC}_{\text{OVR}}$ loss can be obtained directly applying the sigmoid approximation to Eq. (\ref{eq:auc_ovr}).

Considering the computational complexity of the proposed multiclass training objectives, in the case of the $\text{aAUC}_{\text{OVO}}$, it implies computing $c(c-1)$ different aAUCs, resulting in a quadratic growth with the number of classes. On the other hand, in the case of the $\text{aAUC}_{\text{OVR}}$, $c$ different aAUCs have to be computed to obtain the final loss, observing a linear growth with the number of classes.

\section{Experimental setup}
\label{sec:exp_setup}
\subsection{Data description}
In order to demonstrate the capabilities of the proposed multiclass AUC optimisation techniques, we apply them in the context of a 3-class audio segmentation task that aims to separate foreground, background and no music fragments. This task is proposed in the OpenBMAT dataset \cite{openbmat}, being the first of its kind to provide annotations separating foreground and background music. Audio data comes from broadcast domain, with emissions from different countries manually labelled.

The dataset contains around 27 hours of audio from which 16.60\% belongs to foreground music, 34.45\% belongs to background music and 48.94\% belongs to no music. Dataset authors predefined 10 splits, that we use to separate data for training, validation and test purposes: splits zero to seven are used for training, while split eight is reserved for training validation. Split number nine constitutes the testing set. This translates in a total of 22 hours of audio for training, and around 3 different hours for validation and test respectively. All the audio files have been downsampled to 16 kHz and mixed down to a single channel. The OpenBMAT dataset can be downloaded upon request to their respective authors.
\vspace{-0.23cm}

\subsection{Neural network}
In this work, the neural architecture is based on recurrent neural networks (RNN). We stack 2 bidirectional gated recurrent units (GRU) \cite{gru} with 128 neurons each, followed by a linear layer with softmax activation and 3 output neurons. Adam optimiser is used with a learning rate that decays exponentially from $10^{-3}$ to $10^{-4}$ during the 20 epochs that data is presented to the neural network with a minibatch size of 64. Model selection is done by choosing the best performing model in terms of frame classification accuracy using the validation subset. The hyperparameter $\delta$ used to control the slope of the sigmoid has a value of 10 seeking to obtain a shape close to the unit step function, as in \cite{mingoteAUC}. After neural network training, the 3-dimensional scores for the validation subset are extracted and then used to adjust a logistic regression model via a stochastic gradient descent algorithm. This model is used to obtain the final classification labels.

Concerning feature extraction, inspired by our previous experience in audio segmentation tasks \cite{gimeno2018recurrent,recurrent_iberspeech}, we combine a traditional set of perceptual features with some musical theory motivated features. First, 128 log Mel filter bank energies are extracted between 64 Hz and 8kHz. Additionally, they are combined with chroma features \cite{chroma}, extracted using the openSMILE toolkit \cite{openSMILE}. Features are computed every 10 ms using a 25 ms Hamming window. Training and evaluation is done using 3 second sequences (300 frames). Before concatenation, log Mel energies and chroma features are min-max normalised in the range between 0 and 1. 

This setup is fixed for all our experiments, as our main goal is not to evaluate this neural architecture compared to other proposals, but to evaluate whether multiclass AUC optimisation can improve the performance of our audio segmentation system over traditional training objectives.

\subsection{Evaluation metrics}
In order to evaluate the performance of the audio segmentation system, we distinguish two kinds of metrics, those showing information for all the possible thresholds considering the neural network scores, and those showing information for the final decisions made by the classification system. As an aggregate measure of the discrimination of all the classes, we report multiclass AUC in the two variations presented previously: $\text{AUC}_{\text{OVO}}$ and $\text{AUC}_{\text{OVR}}$. We also show the average area under the precision versus recall curve. In order to illustrate each of the class performance, we include the precision versus recall curve per class with the corresponding f1 isocurves. In a similar way as done in MIREX evaluations, we report final classification performance with overall accuracy, and per class precision, recall and f1 \cite{powers2020evaluation}.
\begin{table}
\caption{$\text{AUC}_{\text{OVO}}$, $\text{AUC}_{\text{OVR}}$ and average area under the precision versus recall curve on test data for the audio segmentation systems trained using the proposed multiclass AUC training objectives compared to two variants of cross entropy based training. (Mean $\pm$ standard deviation over 10 different experiments)}
\label{tab:results_AUC}
\small
\setlength{\tabcolsep}{3pt}
   \begin{tabular}{c | m{1.8cm} m{1.8cm} c }
         \toprule
         \parbox[top][1cm][c]{1.3cm}{\textbf{\centering{Training\\objective}}} & \textbf{$\text{\small AUC}_{\text{\scriptsize OVO}}$\small (\%)} & \textbf{$\text{\small AUC}_{\text{\scriptsize OVR}}$\small (\%)} & \parbox[top][1cm][c]{2.1cm}{\centering \textbf{\small Avg. AUC(\%)\\prec vs recall}}\\
         \midrule
         Softmax CE & \multicolumn{1}{c}{81.69$\pm$0.84} & \multicolumn{1}{c}{79.42$\pm$0.69} & 66.05$\pm$0.97\\
         Angular softmax & \multicolumn{1}{c}{80.95$\pm$0.71} & \multicolumn{1}{c}{79.23$\pm$0.65} & 65.89$\pm$0.91\\
         $\text{aAUC}_{\text{OVO}}$ & \multicolumn{1}{c}{83.67$\pm$0.54} & \multicolumn{1}{c}{81.28$\pm$0.61} & 69.55$\pm$0.80 \\
         $\text{aAUC}_{\text{OVR}}$ & \multicolumn{1}{c}{82.46$\pm$0.81} & \multicolumn{1}{c}{80.33$\pm$0.71} & 68.51$\pm$0.90 \\
         \bottomrule
     \end{tabular}
     \vspace{-0.5cm}
\end{table}

All metrics are computed at frame level and without collar on the full test data presented in the previous subsection. Seeking to validate the statistical robustness of our proposal, all our experiments are run 10 different times reporting mean and standard deviation or both of them for the metrics described.

\section{Results}
\label{sec:res}

\begin{figure*}[t]
\centerline{\includegraphics[width=\linewidth]{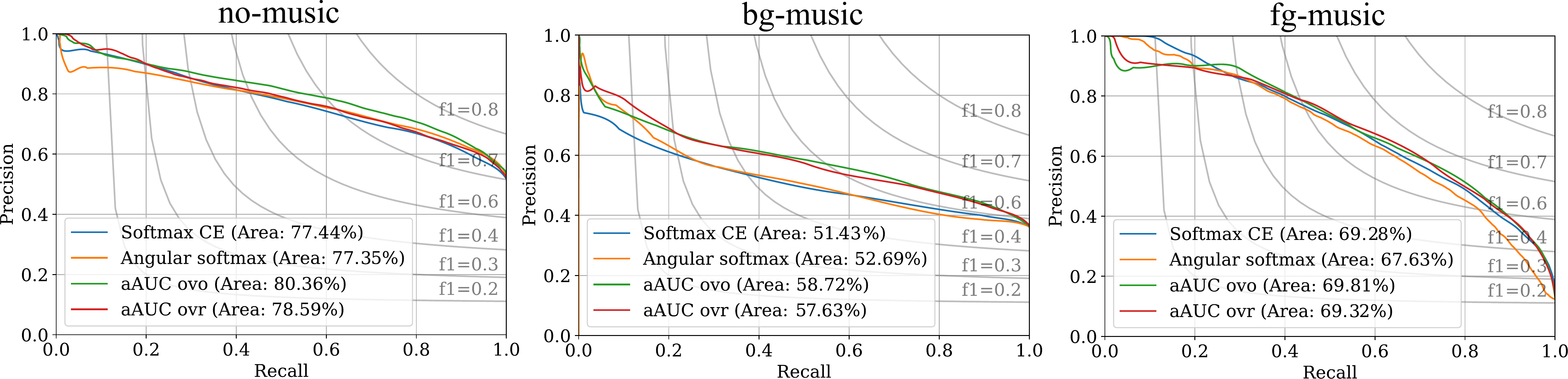}}
\vspace{-0.2cm}
\caption{Precision versus recall curves, f1 isocurves, and area under the precision versus recall curve per class on the test data for the proposed multiclass AUC training objectives compared to two variants of cross entropy based training. (Average curve obtained over 10 different experiments)}
\label{fig:prec_recall}
\vspace{-0.3cm}
\end{figure*}

\begin{table*}[t]
\caption{Final classification overall accuracy(acc), precision(prec), recall(rec) and f1 per class on the test data for the audio segmentation systems trained using the proposed multiclass AUC function losses compared to two variants of cross entropy based training. (Mean $\pm$ standard deviation over 10 different experiments)}
\vspace{-0.15cm}
\label{tab:hard_results}
\small
\setlength{\tabcolsep}{3pt}
\centering
    \begin{tabular}{c|c|c c c|c c c|c c c}
    \toprule
    \multirow{2}{*}{\textbf{Loss function}} &  \multirow{2}{*}{\textbf{acc}} & \multicolumn{3}{c|}{\textbf{no-music}} & \multicolumn{3}{c|}{\textbf{bg-music}} & \multicolumn{3}{c}{\textbf{fg-music}}\\
     &  & \textbf{prec} & \textbf{rec} & \textbf{f1} & \textbf{prec} & \textbf{rec} & \textbf{f1} & \textbf{prec} & \textbf{rec} & \textbf{f1} \\
    \midrule
    Softmax CE & 0.57$\pm$0.03 & 0.76$\pm$0.04& 0.46$\pm$0.05 & 0.57$\pm$0.04 & 0.49$\pm$0.02 & 0.69$\pm$0.06 & 0.57$\pm$0.03 & 0.66$\pm$0.06 & 0.63$\pm$0.04 & 0.63$\pm$0.03\\
    Angular softmax & 0.56$\pm$0.03 & 0.71$\pm$0.04 & 0.48$\pm$0.03 & 0.57$\pm$0.03 & 0.51$\pm$0.03 & 0.63$\pm$0.05 & 0.56$\pm$0.04 & 0.60$\pm$0.04 & 0.64$\pm$0.03 & 0.62$\pm$0.04\\
    $\text{aAUC}_{\text{OVO}}$ & 0.65$\pm$0.02 & 0.81$\pm$0.03& 0.70$\pm$0.03 & 0.75$\pm$0.03 & 0.50$\pm$0.04 & 0.72$\pm$0.05 & 0.59$\pm$0.03 & 0.54$\pm$0.03 & 0.67$\pm$0.04 & 0.60$\pm$0.03\\
    $\text{aAUC}_{\text{OVR}}$ & 0.62$\pm$0.02 & 0.80$\pm$0.03 & 0.51$\pm$0.03 & 0.63$\pm$0.02 & 0.49$\pm$0.02 & 0.70$\pm$0.04 & 0.58$\pm$0.02 & 0.58$\pm$0.02 & 0.71$\pm$0.05 & 0.61$\pm$0.03\\
    \bottomrule
    \end{tabular}
    \vspace{-0.4cm}
\end{table*}

As the starting point of our experimentation, we aim to obtain a baseline system so that further results could be compared. Using the experimental framework described previously, a neural network was trained using a cross entropy loss function. Note that, in this work, cross entropy is implemented as a multiclass loss, also known as softmax cross entropy in the literature. Furthermore, results are also compared with a well known variant of the softmax cross entropy, the angular softmax loss \cite{angular_softmax}. Both systems serve as a comparison point for our proposed multiclass AUC training objectives. Table~\ref{tab:results_AUC} compares the $\text{AUC}_{\text{OVO}}$, $\text{AUC}_{\text{OVR}}$ and average area under the precision versus recall curve on test data for our proposed multiclass AUC loss functions and the two cross entropy training variants. It must be noted that both $\text{aAUC}_{\text{OVO}}$ and $\text{aAUC}_{\text{OVR}}$ provide an improvement compared to softmax and angular softmax training in all the metrics presented. This is most noticeable in the OVO approach that yields an approximate 2.5\% average relative improvement in both AUC multiclass metrics and an approximate 5\% average relative improvement in area under the precision versus recall curve when compared to softmax CE training.

Figure \ref{fig:prec_recall} summarises the neural network scores performance per class for different thresholds comparing our proposed AUC multiclass loss functions to the two cross entropy variations presented, showing precision versus recall curves and the corresponding f1 isocurves. In general, we can see that both of our proposals achieve an improvement in the classification of no music and background music classes, without a significant drop in performance in the foreground music class. The improvement in the background music class is consistent between the OVO and OVR approach (around 14\% average relative improvement compared to cross entropy training in area under the precision versus recall curve). However, the OVO approach is able to obtain better results on the no music class (around 4\% average relative improvement compared to cross entropy training in area under the precision versus recall curve), achieving a better overall result and matching the metrics shown in Table \ref{tab:results_AUC}. It can be observed again that, when comparing softmax CE and angular softmax losses, the last one yields similar results to softmax CE training with a minor overall drop in performance.

Finally, Table \ref{tab:hard_results} shows the results obtained by the final classification labels in terms of overall accuracy and per class precision, recall and f1 for our proposed multiclass AUC training objectives compared to both cross entropy variants. In this case, the AUC optimisation techniques derive in an average relative improvement of 14.03\% and 8.77\% respectively for the OVO and OVR approach, compared to softmax CE in terms of overall accuracy. Again, the OVO approach seems to outperform the OVR approach. This suggests that, in general terms, the combination of pairs of classes performed in the OVO solution is more powerful than the binary transformation performed in the OVR solution. It can be observed that this boost in performance comes mainly from the increment in the f1 metric for the no music class that rises from an average value of 0.57 in the softmax CE training to 0.75 in the $\text{aAUC}_{\text{OVO}}$ training. Angular softmax shows a similar trend as the one observed in the previous results, with a performance slightly below softmax CE training and underperforming our proposed multiclass AUC loss functions. 

\section{Conclusion}
\label{sec:conc}
In this paper, we have introduced a generalisation of the AUC optimisation framework that can be applied to an arbitrary number of classes. Derived from the presented multiclass extensions of the AUC metric, we introduce two training objectives based on a one-versus-one and a one-versus-rest approach respectively. We evaluate our proposal in a 3 class audio segmentation task that aims to separate foreground music, background music and no music using an experimental setup based on recurrent neural networks with a limited training data set consisting only of around 20 hours of audio. Despite the validity of the theoretical framework presented, further research would be needed to explore the possible effects of using a larger number of classes on the performance of the proposed training objectives.

Experimental results suggest that the multiclass AUC optimisation outperforms traditional training objectives such as softmax cross entropy or angular softmax, under the circumstances of a limited training dataset being used. In particular, for this task, we report an average relative improvement close to 14\% in terms of overall accuracy using our proposed $\text{aAUC}_{\text{OVO}}$ training criterion. Comparing the two different approaches considered, the OVO approach outperforms the OVR approach in all of our experiments. This fact suggests that the combination of pairs of classes is a more robust training criterion than the use of one-versus-rest binarisation solutions.

\bibliographystyle{IEEEtran}
\bibliography{references}

\end{document}